\documentclass[10pt,onecolumn,superscriptaddress,nofootinbib,floatfix,amsmath,amssymb]{revtex4}

\usepackage[utf8]{inputenc}
\usepackage{amsmath}
\usepackage{slashed}
\usepackage{graphicx}
\usepackage[caption=false]{subfig}
\usepackage{xcolor}
\usepackage[pdftex]{hyperref}
\begin{document}

\title{ Deformation of  Nanowires and Nanotubes}
\author{Aatif Kaisar Khan}
\affiliation{Department of Metallurgical and Materials Engineering, NIT Srinagar, Kashmir 190006 India}
\author{Salman Sajad Wani}
	\affiliation{Department of Physics and Engineering, Istanbul, Technical University, 34469, Istanbul, Turkey}
\affiliation{Canadian Quantum Research Center 204-3002, 32 Ave Vernon, BC V1T 2L7 Canada}
\author{  Aasiya Shaikh }
 \affiliation{Design and Manufacturing Technology Division, Raja Ramanna Centre for Advanced Technology, Indore-452013, Madhya Pradesh, India}
\author{Yas Yamin}
\affiliation{Irving K. Barber School of Arts and Sciences, University of British Columbia Okanagan Campus, Kelowna, BC V1V1V7, Canada}
\author{Naveed Ahmad Shah}
\affiliation{Department of Physics, Jamia Millia Islamia, New Delhi 110025 India}
 \author{Yermek O. Aitenov}
\affiliation{Department of Genetics, University of Cambridge, Downing Street, Cambridge CB2 3EH, United Kingdom}
\author{Mir Faizal}
\affiliation{Canadian Quantum Research Center 204-3002, 32 Ave Vernon, BC V1T 2L7 Canada}
\affiliation{Irving K. Barber School of Arts and Sciences, University of British Columbia Okanagan Campus, Kelowna, BC V1V1V7, Canada}
\author{Suhail Lone}
\affiliation{Department of Physics and Astrophysics, University of Delhi, Delhi 110007, India}

\begin{abstract}
In this article, we have investigated the consequences of the next to the leading order correction to the effective field theory of  nanostructures. This has been done by analyzing the effects of deformed Heisenberg algebra on  nanowires and nanotubes. We  first deform the Schrodinger equation with cylindrical topology.  Then specific solutions to the deformed Schrodinger equation with different boundary conditions are studied. These deformed solutions are  used to investigate  the consequences  of the deformation on the energy of nanowires and nanotubes. This deformation can be detected by connecting such nanostructures  to  ferromagnets, and testing the  current-voltage relation for such junctions.
\end{abstract}

\maketitle

\section{Introduction} 
Even though the condensed matter systems are made of discrete atoms, the effects of discreteness do not show up at the leading order. However,  due to  the next to nearest atom hopping the effective field theory of condensed matter systems gets deformed    \cite{gupa}. This deformation can be viewed as a deformation caused by a modification of the uncertainty principle to a generalized uncertainty principle (GUP)  \cite{ml12, ml21}.  This modification of the uncertainty principle deforms the Heisenberg algebra. The deformed Heisenberg algebra can also be obtained  using next to the leading order correction  the effective field theory representing such  systems \cite{gupq}. Thus, the GUP deformation can be viewed as a next to the leading order correction to the effective field theory of a condensed matter system.  This is because the  deformed theory captures the behavior of the next to nearest atom hopping for certain condensed matter systems. In fact, the GUP deformation is the first order correction to the continuum approximation and occurs due to the discrete nature of condensed matter systems. Furthermore, the tight-binding model has been used to show that beyond the linear regime the energy-momentum  dispersion relation gets modified \cite{gup12, gup14}. Now such modified energy-momentum  dispersion relations are known to be related to GUP \cite{gup15, gup16}. Thus, it is expected that beyond the linear regime the condensed matter systems will be represented by GUP deformed structures. For such materials, it has been explicitly demonstrated that the generalized Dirac structure which emerges beyond the linear regime is the GUP deformation of the usual Dirac structure \cite{gupb}.  It is also possible to obtain GUP deformation of the effective theory describing condensed matter systems using spin-orbit interaction \cite{spin}. 
It may be noted that GUP deformation occurs due to an intrinsic minimal length in the system \cite{ml12, ml21}. This minimal length is realized in atomic systems as inter-atomic distance, and so the GUP deformation of condensed matter systems occurs due to this intrinsic length \cite{gupb}. 

As the GUP deformation occurs due to the next to the leading order correction to the effective field theory of a system \cite{gupq}, it has been used to analyze quantum gravitational effects \cite{s16}. An intrinsic minimal length which occur due to such  quantum gravitational effects can be detected using ultra-precise measurements of GUP deformed condensed matter systems  \cite{ml12, ml21}. The  ultra-precise measurements of Landau levels and Lamb shift have been used to put a bound on such a minimal length \cite{ml15}. This has been done by investigating the GUP deformation of Landau levels and Lamb shift and then comparing such deformation with observational data. It has also been suggested that a GUP deformation of an  opto-mechanical setup can be used to measure this minimal length  \cite{ml14}. It has been demonstrated that a general GUP deformation can lead to Discrete space \cite{space}. This has been extended to relativistic quantum mechanics \cite{space1}. 
Thus, there are two different motivations to analyze the GUP deformation of condensed matter systems: the first one is due to the discrete nature of atomic systems, and the second one is due to the possibility of detecting exotic quantum gravitational effects through ultra-precise measurements. Even though the physical motivations are totally different, the deformation of the Heisenberg algebra produced from both of them is the GUP deformation. Hence, here we analyzed the GUP deformation of nanotubes and nanowires, without specifying the origin of such a deformation.  

It may be noted that the occurrence of  a minimal length is almost a universal feature of different approaches to quantum gravity  \cite{s16}. The existence of such a minimal length can even be argued using  black hole physics. The  energy needed to make Trans-Planckian measurements is more than the energy needed to form a mini black hole. Thus, any attempt to make Trans-Planckian measurements will result in the formation of such a mini black hole, which will in turn prevent such measurements    \cite{z4,z5}. Now as any theory of quantum gravity has to be consistent with the black hole physics, any such theory should have an intrinsic minimal length associated with it. This minimal length even occurs in theories where it is least expected, such as asymptotically safe gravity \cite{asgr} and conformally quantized quantum gravity \cite{cqqg}.   A minimal length in the form of polymer length also occurs in loop quantum gravity \cite{z1, po12, po14, 12p}.    
Even though it can be argued that a minimal length of the order of Planck should occur in any theory of quantum gravity \cite{s16}, it is possible for this minimal length to be much larger than the Planck length. In fact, such a minimal length much larger than Planck length has been proposed to exist in string theory     \cite{z2,zasaqsw}. This minimum measurable length occurs in perturbative string theory, as the length of the  fundamental string   \cite{cscds,2z}. This is  because such a  fundamental string  is the smallest probe available in such a  theory.   It has been observed that due to T-duality, a minimum length also occurs even after taking into account  non-perturbative objects  in string theory \cite{s16, s18}.  It has been demonstrated that an effective path integral constructed by neglecting all the string oscillatory modes has an intrinsic minimal length associated with it \cite{green1, green2}.
  
Even though interesting work has been done on the GUP deformation of various quantum systems, the GUP deformation of a quantum system with cylindrical topology has not been investigated. However, it is important to analyze the  GUP deformation of systems with cylindrical topology as various nanostructures have this topology. A  nanowire is one such nanostructure with cylindrical topology, with a diameter in the range of $1-100 nm$ and a length ranging from few hundreds of nanometers to millimeters \cite{wire,wire0}. Nanowires can be produced by either  top-down or bottom up approaches.  Electron beam lithography is used in top-down approach    \cite{wire1, wire2}, whereas  atoms are combining form a nanostructure in the bottom-up approach  \cite{wire4}. It is also possible  to produce nanowires using  DNA strands as scaffolds \cite{dna1, dna2}. Here the imino proton of each base pair is substituted with  a metal ion, and this alters the electronic properties of the system.  
Another nanostructure with   cylindrical topology is a nanotube \cite{nano1, nano2}. 
A nanotube is a basically a hollow   cylindrical  tube    with a  nanoscale  diameter in the range of $1-50 nm$. The study of nanotubes stated with the discovery of carbon nanotubes \cite{1}. Due to their interesting properties  nanotubes  have   been throughly studied \cite{2,3,4,5,6,7}. Nanotubes have been synthesized  from other materials such as  and used for different applications such as $TiO_2$, due to their    interesting optical and electronic  properties  \cite{8a, 8, 9, 10, 11, 12}.  
These     nanotubes  are usually produced using different methods like  hydrothermal method \cite{13, 14, 15} and anodization method \cite{15a}.  It is also possible to use DNA to obtain nanotubes \cite{dna4, dna5, dna6, dna7}. These DNA nanotubes can be analyzed  as  semiflexible polymers with tunable bond strength. 
As nanowires and nanotubes have been obtained in lab, there is a strong motivation to  analyze the modification to their quantum mechanics from GUP deformation. Thus, we analyze the modification to a quantum system with  cylindrical topology in this paper. It may be noted that the Schrodinger equation for cylindrical geometry has been used to analyze the influence of  geometrical defects on the energy gap and charge distribution \cite{1a}. However, here we  explicitly derive an analytical solution with cylindrical topology, and  modify it  using  GUP corrections.  We  use this GUP deformed solution for investigating the effect of GUP deformation on the energy in nanowires and nanotubes. 

\section{Deformation}
The GUP deformation of condensed matter systems is interesting as it can occur  as a first order correction to the continuum approximation of discrete atomic system \cite{gupa, gupb}. It is also interesting as ultra-precise measurements of condensed matter systems  can be used to measure quantum gravitational effects, which also cause GUP deformation of these condensed matter systems   \cite{ml12, ml21}.
Even though there are different physical    motivation for the GUP deformation,  the deformation in  both these cases depends on a free parameter $\alpha$. So, we apply the GUP deformation to nanowires and nanotubes, without specifying its origin. Thus, we first review a general GUP deformation \cite{ml15} of the  Heisenberg uncertainty principle $\Delta x \Delta p \geq \hbar/2$. This GUP deformation deforms the Heisenberg uncertainty principle to    \cite{ml15}
\begin{equation}
	\begin{split}
		\Delta x \Delta p &\geq \frac{\hbar}{2}[1 - 2 \alpha \langle p \rangle  + 4 \alpha^2 \langle p^2 \rangle]  \\
		&\geq \frac{\hbar}{2}\left[1 + \left( \frac{\alpha}{\sqrt{\langle p^2 \rangle}} + 4 \alpha^2 \right) \Delta p^2 + 4 \alpha^2 \langle p \rangle^2 - 2 \alpha \sqrt{\langle p^2 \rangle}\right]~
	\end{split}
\end{equation}
As the uncertainty principle  is closely related to the Heisenberg algebra, a deformation of the uncertainty principle results in a deformation of the Heisenberg algebra.  The   deformation  of the Heisenberg algebra consistent with the generalized uncertainty principle can be written as \cite{space, space1}
\begin{equation}
	[x_i,p_j] = i \hbar \left[ \delta_{ij} - \alpha \left(p \delta_{ij} + \frac{p_i p_j}{p}\right) + \alpha^2 (p^2 \delta_{ij} + 3 p_i p_j) \right]~ \label{eqn:GUP}
\end{equation}
The  position commutes with position, and momentum with momentum, in this deformation,  
$	[x_i, x_j] = 0 = [p_i,p_j]~.$
Due to this property the Jacobi identity is satisfied  \cite{ml15}. The deformation of the Heisenberg algebra causes the deformation of the coordinate representation of the momentum operator.  However, it is possible to express the deformed momentum operator in terms of the usual momentum operator. 
Thus,  the  deformation  modifies the momentum operator $  {p}^2$   to $   {p}^2  (1-2\alpha   {p  }- 5\alpha^2  {p}^2) $, 
where ${p}$ is the usual momentum with the usual coordinate representation.  It is possible to investigate the effects of this GUP deformation on angular momentum.  Here,  the eigenvalue  $l^2 $ gets modified  to $l^2(1-\langle c \rangle^2)$, with $c = \alpha p - \alpha^2 p^2$ \cite{gup1, gup2}. 

As a nanowire has a cylindrical topology,  the Hamiltonian for the nanowire consists of a two-dimensional circular part and a  part representing the length of the nanowire.  It may be noted that the additional terms ${\hbar^2}/{8m\rho^2}$ occurs due to the  two dimensions circular part, and this term is  important to express the Hamiltonian for a system with  cylindrical topology   \cite{hami}.
So,   the Hamiltonian for a nanowire with  cylindrical topology can be written as  \cite{hami}
\begin{equation}
  {H}\psi(\rho,\phi,z)=\bigg(\frac{  {P}^2_\rho}{2m}-\frac{\hbar^2}{8m\rho^2}+\frac{  {L}_z^2}{2m\rho^2}+\frac{  {P^2_z}}{2m}+V(\rho,z)\bigg)\psi(\rho,\phi,z)
\label{a1}
\end{equation}
Here,  we have defined $  {P}^2_\rho$ and $L_z^2$ as 
\begin{eqnarray}
  {P}^2_\rho=\frac{-\hbar^2}{2m}\bigg( \frac{\partial^2}{\partial \rho^2}+\frac{1}{\rho}\frac{\partial}{\partial \rho}-\frac{1}{4\rho} \bigg),
\label{a2}
&&
L_z^2=-\hbar^2\frac{\partial^2}{\partial\phi^2}
\label{a3}
\end{eqnarray}
 As a nanotube is formed by a very thin layer of material, we can assume it has a constant radius $\rho_0$. 
So, the  Hamiltonian for a nanotube can be represented  as 
\begin{equation}
  {H}\psi(\phi,z)=\frac{-\hbar^2}{2m}  \bigg( \frac{\partial^2}{\partial z^2}+\frac{1}{\rho_0^2}\frac{\partial}{\partial \phi}\bigg)\psi(\phi,z)
\label{a22.1}
\end{equation}
We can deform the $  {P^2_z}$ for both the nanowires and nanotubes  using standard GUP deformation \cite{ml15}. The angular part of both these nanostructures can be  deformed using the standard treatment of angular momentum in GUP deformed theories 
 \cite{gup1, gup2}. This can be used to investigate the deformation of the radial part of the Schrodinger equation. Thus, we can use the standard formalism of GUP deformation to analyze its effects  on nanowires and nanotubes. 

\section{   Nanowires }
In this section, we  analyze the  GUP deformation in  nanowires, and  obtain  the solutions to the GUP deformed   Schrodinger equation with cylindrical topology.  
It is proposed that the  solution has the following form:
\begin{equation}
\psi(\rho,\phi,z)=Z(z)\Phi(\phi)R(\rho)
\label{a4}
\end{equation}
Now, having separated the variables with our choice of the ansatz, we separately investigate the axial ($Z(z)$), radial ($R(\rho)$) and the angular ($\Phi(\phi)$) parts of the solution.  

\begin{figure}[htbp]
\begin{center}
\includegraphics[scale=0.75]{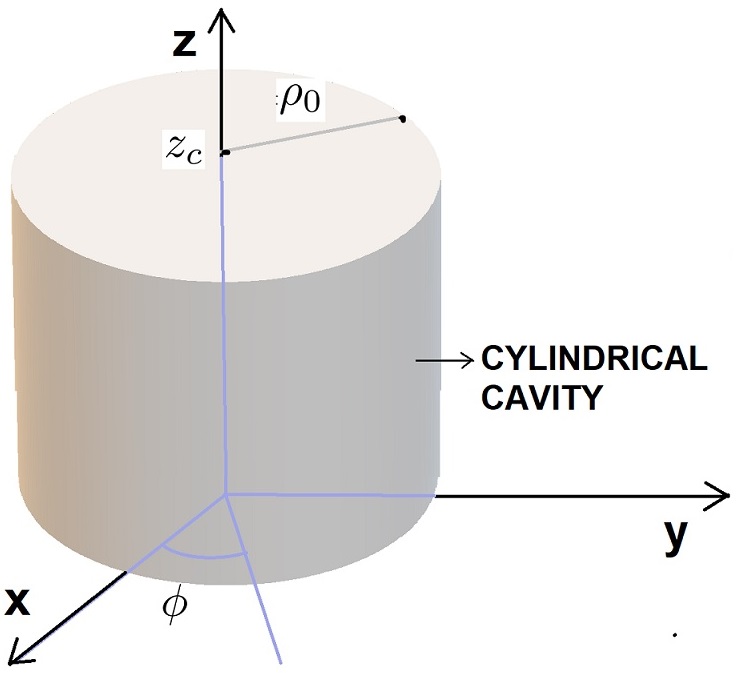}\\    
\end{center}
\caption{ Nanowire  }
\end{figure}

To  obtain the solution for $Z(z)$,  the eigenvalue equation  is written as  \cite{gup1, gup2}
\begin{equation}
  {P}_{z}^2(1-2\alpha   {P_{z}}- 5\alpha^2  {P}_{z}^2)Z(z)=2m\lambda Z(z),
\label{a5}
\end{equation}
where $ {P}_{z}=-i\hbar{\partial}/{\partial z}$. 
Neglecting the higher  powers of $\alpha$,  and the solving   above equation, the following expression is obtained  
\begin{equation}
Z(z)=C_1e^{ik_{1z}z}+C_2e^{-ik_{2z}z}+C_3e^{\frac{i}{2\alpha\hbar}z}
\label{a7}
\end{equation}
where $k_{1z}=k_z(1+\alpha\hbar k_z)$ and $k_{2z}=k_z(1-\alpha\hbar k_z)$ are 
the deformed wave-vectors, with $ k_z=\sqrt{ {2m\lambda}/{\hbar^2}}$. 
Now    the solution of $Z(z)$ for $\alpha=0$, can be expressed as  
\begin{equation}
Z(z)=C_1e^{ik_z z}+C_2e^{-ik_z z}
\label{a9}
\end{equation}
The original  solution can be obtained by neglecting the deformation parameter, i.e., by setting $\alpha = 0$. The solution without the third term can be understood as a perturbative solution, with $\alpha$ as the small perturbative parameter. However,  the third term in  Eq. \eqref{a7} is produced due to non-perturbative effects. It represents   the quantization of background geometry. From Eqs. \eqref{a7} and \eqref{a9}, we conclude that $\lim_{\alpha\to 0} |C_3|=0$. Here,  $C_1$ can be made real by neglecting the phase from $Z(z)$. The boundary conditions are given by $Z(z)=0 $ for $ z=0, z_c$, where $0$ and $z_c$ are the axial coordinates of two ends of the nanowire. 
The boundary condition  $Z(0)=0$ can be used   to eliminate $C_2$
\begin{eqnarray}
\psi(z) &=&  2i \ C_1 \ \sin(k_z  z)+ C_3 \ (-e^{ik_z  z}  + e^{\frac{i}{2\alpha\hbar}z}).  \nonumber \\
&& -\alpha\hbar \ k_z^2 \
 z (iC_3 \ e^{-ik_zz}+2C_1 \ \sin(kz))
\label{a10.2}
\end{eqnarray}
The second boundary condition; $Z(z_c)=0$, is used to obtain 
\begin{eqnarray}
2i \ C_1 \ \sin(k_z \ z_c) &=& |C_3| \ (e^{-i(k_z \ z_c+\theta_c)} - \ e^{i(\frac{z_c}{2\alpha\hbar}-\theta_c)}) \nonumber \\
&&+\alpha \hbar \ k_z^2 \ z_c(i|C_3| \ e^{-i(k_z \ z_c+\theta_c)}+2C_1 \ \sin(k_z \ z_c)),
\label{a10.3}
\end{eqnarray}
where $C_3 = |C_3|e^{-i\theta_c}$.
From the above equation, it can be seen that as $\alpha\to0$, both the terms on the right tend to zero. Therefore, we infer from this  equation,  that  $k_zz_c=n\pi$, where $n\in\mathbb{Z}$. So, as $\alpha\neq0$, an addition to the phase on the left hand side of the above equation is obtained  i.e., $k_zz_c=n\pi+\delta_c$, where $\lim_{\alpha\to0}\delta_c=0$. Hence, from this result and the expression for $C_3$, we can conclude that second term in right hand side of Eq. \eqref{a10.3} drops off more rapidly than its first term,  and therefore can be neglected. Thus, we can write 
\begin{equation}
2iC_1\sin(k_zz_c)=|C_3|(e^{-i(k_zz_c+\theta_c)}-e^{i(\frac{z_c}{2\alpha\hbar}-\theta_c)}).
\label{a10.4.1}
\end{equation}
Now, knowing that $C_1\in\mathbb{R}$, and equating the real parts of above equation, we obtain
\begin{eqnarray}
\cos \left(\frac{z_c}{2\alpha\hbar}-\theta_c\right)&-&\cos(n\pi+\theta_c+\delta_c)  = \\
&& 2\sin\bigg( \frac{\frac{z_c}{2\alpha\hbar}+n\pi+\delta_c}{2} \bigg) \sin\bigg( \frac{\frac{z_c}{2\alpha\hbar}-n\pi-2\theta_c-\delta_c}{2} \bigg) = 0.
\label{a10.6}
\end{eqnarray}
Further, from the above equation,  the two conditions are derived 
\begin{equation}
\frac{z_c}{2\alpha\hbar}=(2q+n)\pi+2\theta_c +\delta_c \ \ , \ \ \frac{z_c}{2\alpha\hbar}=(2q-n)\pi -\delta_c 
\label{a10.7}
\end{equation}
where, $2q\pm n$ can only have values such that $z_c$ is positive (with the additional phase to $k_z z_c$ from deformation being $ \delta_c $).   

\begin{figure}[htbp]
\begin{center}
 \includegraphics[scale=0.75]{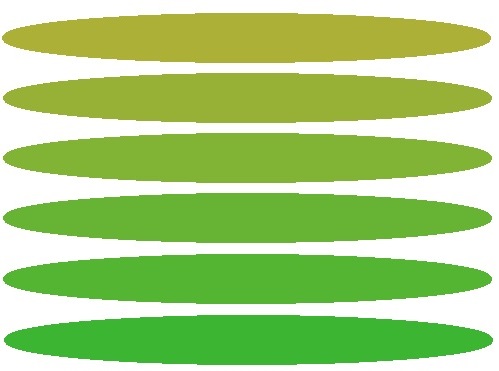}   
\end{center}
\caption{Perceived Disc Structure of Space}
\label{fig:discs}
\end{figure}

These equations represents the discretization for the height of the cylinder.   The existence of a particle inside this cylinder is subjected to this configuration of the cylinder. This shows the discrete nature of space in a three dimensions with space under this specific potential essentially being composed of discs aligned in the $z$-direction of the cylinder (see Fig. \ref{fig:discs}). This quantization of space is very similar to what has been calculated for {particles} in a one-dimensional box \cite{ml15}.

Using the angular part of the ansatz \eqref{a4} in the  Schrodinger  Eq. \eqref{a1}, for $\Phi(\phi)$,  we get \cite{gup1, gup2}
\begin{equation}
  {L}_z^2\Phi(\phi)=l^2(1-\langle c \rangle^2)\Phi(\phi)
\label{a11}
\end{equation}
Here, $c$ is defined as $c = \alpha p - \alpha^2 p^2$. 
Again for the angular case 
the solution is given by $\Phi(\phi)=C_4e^{ik_\phi \phi}+C_5e^{-ik_\phi \phi},$
where $k_\phi=({l^2}/{\hbar^2})({(1-\langle c \rangle^2)}/{(1-2\alpha \rho_0)})$. Using the boundary condition $\Phi(0)=\Phi(2\pi)$, we get
\begin{equation}
\Phi(\phi)=C_4(e^{ik_\phi \phi}+e^{-ik_\phi (2\pi-\phi)})
\label{a16}
\end{equation} 
\begin{figure}[htbp]
\begin{center}
\includegraphics[scale=0.70]{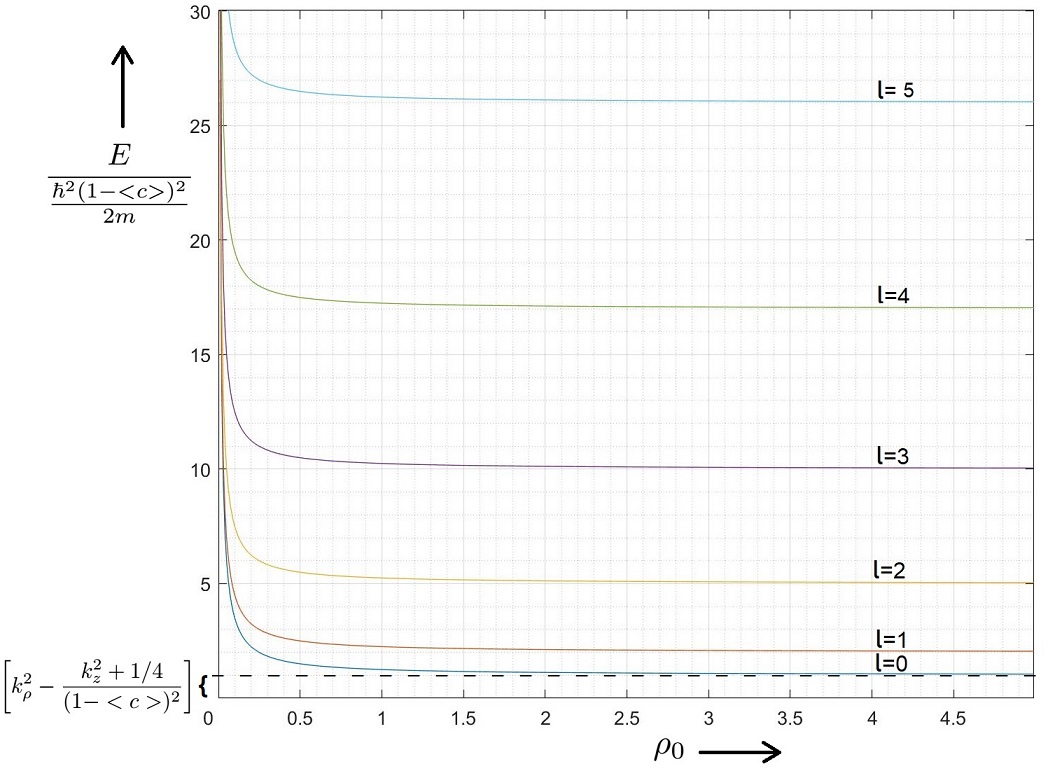}\\   
\end{center}
\caption{Relation Between Energy and Radius }
\label{fig:plot2}
\end{figure}

The Schrodinger equation for the radial part can be written using Eqs. \eqref{a4} and \eqref{a1}, and is given by \cite{gup1, gup2}
\begin{equation}
\frac{-\hbar^2}{2m}\bigg( \frac{\partial^2}{\partial \rho^2}+\frac{1}{\rho}\frac{\partial}{\partial \rho}-\frac{1}{4\rho_0} \bigg)(1-\langle c \rangle)^2R(\rho)=\frac{\hbar^2}{2m}\bigg( \frac{1}{4}-l^2(1-\langle c \rangle)^2\bigg)R(\rho)+(E+\lambda)R(\rho)
\label{a17}
\end{equation}
The solution is of the form
\begin{equation}
R(\rho)=C_6J_0(k_\rho\rho)+C_7Y_0(k_\rho\rho)
\label{a18}
\end{equation}
where, $J_0$ and $Y_0$ represent Bessel functions. The form of the radial deforms the  wave-vectors as 
\begin{equation}
k_\rho^2=\frac{2m}{\hbar^2}\frac{(E+\lambda)}{(1-\langle c \rangle)^2}+\frac{1}{4(1-\langle c \rangle)^2}-\frac{1}{4\rho_0}-l^2
\label{a19}
\end{equation}
Hence using the expressions for $k_{1z},k_{2z}$ in the above equation,
$
E=({\hbar^2}/{2m}) ( (1-\langle c \rangle)^2(k_\rho^2+({4\rho_0})^{-1}+l^2)-{4}^{-1} )-({\hbar^2k_z^2}/{2m}).
\label{a20}
$
Since, $c$ is directly proportional to $\alpha$, we can write 
\begin{equation}
E= \left(\frac{\hbar^2(1- \langle c \rangle)^2}{2m}\right)^{-1} \left(\bigg( k_\rho^2-\frac{k_z^2+1/4}{(1-\langle c \rangle)^2} \bigg)+(\frac{1}{4\rho_0}+l^2)\right)
\label{a20.2}
\end{equation}
The effect of $\rho_0$ and $l$ on $E$  has been plotted in Fig. \ref{fig:plot2}. 
Using the boundary condition $R(\rho_0)=0$ in Eq. \eqref{a18}, 
\begin{equation}
R(\rho)=C_6\bigg(J_0(k_\rho\rho)-\frac{J_0(k_\rho\rho_0)}{Y_0(k_\rho\rho_0)} Y_0(k_\rho\rho)\bigg).
\label{a22}
\end{equation}
This is the solution to the GUP deformed $R(\rho)$ for nanowires. 
\section{ Nanotube}
\begin{figure}[h!]
\begin{center}
\includegraphics[scale=0.75]{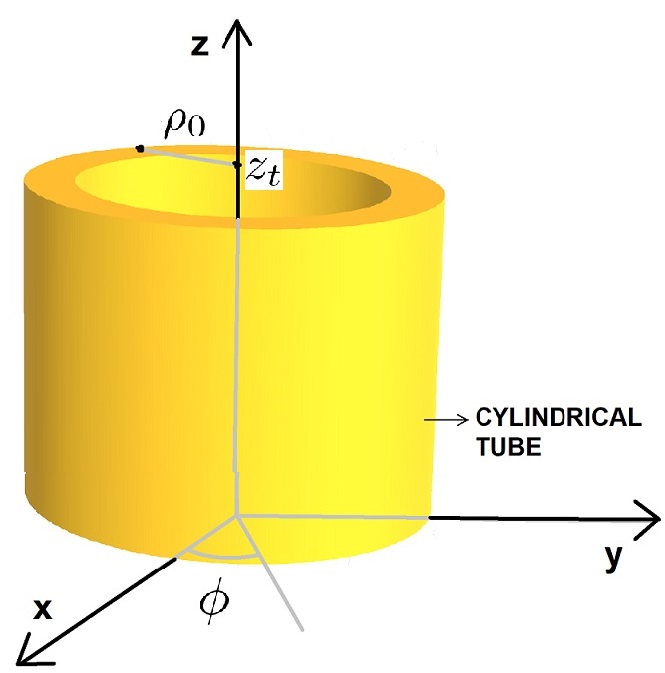}\\
\end{center}
\caption{Nanotube}
\label{fig:nanotube}
\end{figure}

\begin{figure}[htbp]
\includegraphics[scale=0.70]{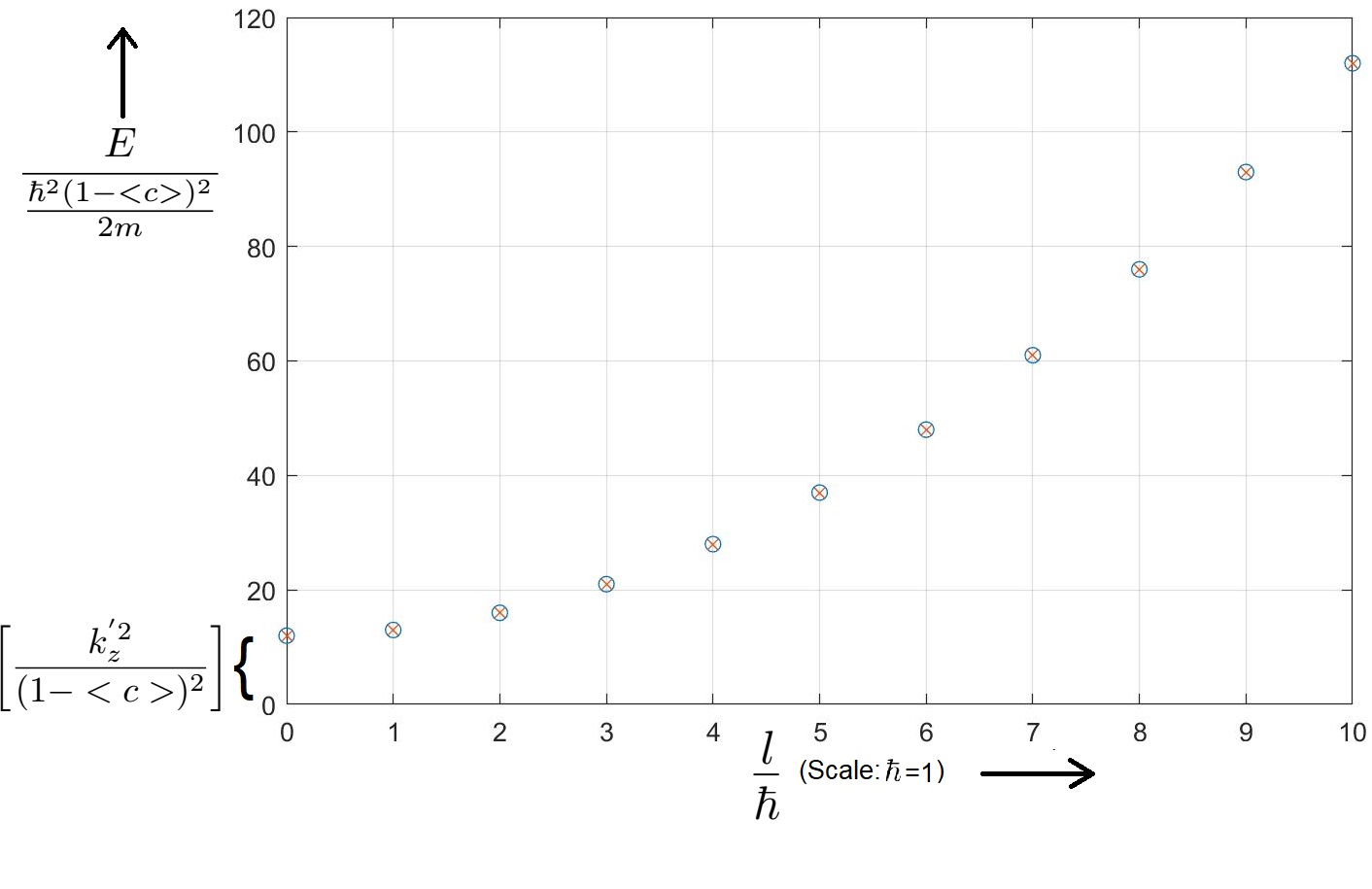}\\
\caption{Relation Between Energy and $l$ for  nanotube}
\label{fig:plot}
\end{figure}

In this section, we  analyze the  GUP deformation in  nanotubes. We  again obtain  the solutions to the GUP deformed   Schrodinger equation with cylindrical topology. However, here we analyze the solution to the  Schrodinger equation with a hollow cylinder (see Fig. \ref{fig:nanotube}). 
We propose the following  ansatz for the solution to GUP deformed nanotubes   
\begin{equation}
\psi(\phi,z)=\Phi(\phi)Z(z)
\label{a22.2}
\end{equation}
As in the case of the nanowires, we can write the Schrodinger equation for the axial part of the wave function using  \cite{gup1, gup2}
\begin{equation}
  {P}_{z}^2(1-2\alpha   {P_{z}}- 5\alpha^2  {P}_{z}^2)Z(z)=2m\bigg( E-\frac{l^2(1-\langle c \rangle)^2}{2m} \bigg) Z(z),
\label{a23}
\end{equation}
where $   {P}_{z}=-i\hbar{\partial}/{\partial z}$. Here, 
  we get a solution with three terms
\begin{equation}
Z(z)=S_1e^{ik'_{1z}z}+S_2e^{-ik'_{2z}z}+S_3e^{\frac{i}{2\alpha\hbar}z}
\label{a25}
\end{equation}
where $k'_{1z}=k'_{z} (1+\alpha\hbar k'_z)$, $k'_{2z}=k'_z(1-\alpha\hbar k'_z)$ and $k'^{2}_{z}= ({{2m}/{\hbar^2})(E- ({l^2(1-\langle c \rangle)^2}/{2m}))}$.
The energy eigenvalues are given by
$ E=\hbar^2k_z^{'2}/2m+l^2(1-\langle c \rangle)^2/2m$. 
Using  $\alpha \propto c$, we can write  $E=({\hbar^2}/{2m})(k_z^{'2}+{l^2}{\hbar^{-2}})-\langle c \rangle({l^2}/{m})$.
From the above expression,  the energy can be expressed as 
\begin{equation}
E= \left(\frac{\hbar^2(1-\langle c \rangle)^2}{2m}\right)^{-1} \bigg[ \frac{k_z^{'2}}{(1-\langle c \rangle)^2} \bigg]+\bigg(\frac{l}{\hbar}\bigg)^2
\label{a26.3}
\end{equation}
The dependence of energy, $E$, on  $l$ has been plotted in  Fig.\ref{fig:plot}. 

Here,    a solution of the form $Z(z)=S_1e^{ik'_z z}+S_2e^{-ik'_z z}$ is obtained for $\alpha=0$. After comparing this solution with the case where we have  a non-zero value to the parameter $\alpha$ i.e. in Eq. \eqref{a25}, we infer that in the limit $\alpha\to 0$, $|S_3|=0$.

Here,  $S_1$ is made real by  neglecting the phase  from $Z(z)$. The boundary conditions are given by $Z(z)=0 $ and  for $ z=0, z_t$.
Therefore, using $Z(0)=0$ to eliminate $S_2$ in Eq. \eqref{a25} and  using $Z(z_t)=0$, the following expression is obtained 
\begin{eqnarray}
2i \ S_1\sin(k_z \ z_t) &=& |S_3| \ (e^{-i(k_z \ z_t+\theta_t)}-e^{i(\frac{z_t}{2\alpha\hbar}-\theta_t)})  \nonumber \\
&+& \alpha \hbar \ k_z^2 z_t \ (i|S_3 \ |e^{-i(k_z \ z_t+\theta_t)}+2S_1 \ \sin(k_zz_t)),
\label{a29}
\end{eqnarray}
where $S_3=|S_3|e^{-i\theta_t}$.
Here, Eq.  \eqref{a29} is similar to Eq. \eqref{a10.3},  and hence it can be again seen that as $\alpha\to0$, and the right hand side tends to  zero. Hence, in  the left hand side  $k_zz_t=n\pi$, where $n\in\mathbb{Z}$.  As $\alpha\neq0$, we get a condition on the left hand side,  i.e.,  $k_zz_t=n\pi+\delta_t$ (where, $\lim_{\alpha\to0}\delta_t=0$). So, using this and the properties of  $S_3$,  that second term in right hand side of Eq. \eqref{a29} has a sharp drop,  and can be neglected. Thus, we can write 
\begin{equation}
2iS_1\sin(k_zz_t)=|S_3|(e^{-i(k_zz_t+\theta_t)}-e^{i(\frac{z_t}{2\alpha\hbar}-\theta_t)})
\label{a31}
\end{equation}
By following the method used to obtain Eq.  \eqref{a10.7},  the following two conditions are obtained 
\begin{equation}
\frac{z_t}{2\alpha\hbar}=(2q+n)\pi+2\theta_t +\delta_t, \ \ \ \ \    \frac{z_t}{2\alpha\hbar}=(2q-n)\pi -\delta_t
\end{equation}
where $2q\pm n$ can only have values such that $z_t$ is positive (with the additional phase to $k_z z_t$ from deformation being $ \delta_t $).
These equations represent the discretization for the length of the nanotube. Thus, any the particle inside this nanotube can exist only at the specific lengths of the tube. It can be interpreted as the discreteness in the background geometry of  the two-dimensional curved surface representing such a nanotube. The space under this specific potential seems to be composed of rings aligned in the $z$-direction of the cylinder. This quantization of space is similar to what has been discussed for a cylindrical cavity.  
Solution for $\Phi(\phi) = S_4(e^{ik_\phi \phi}+e^{-ik_\phi (2\pi-\phi)})$ has the same form as Eq. \eqref{a16}, as its differential equation and boundary condition remains the same.
Thus, the behavior of $\Phi(\phi)$ for nanotubes is similar to their behavior in nanowires. 
\begin{figure}[htbp]
\begin{center}
\includegraphics[scale=0.75]{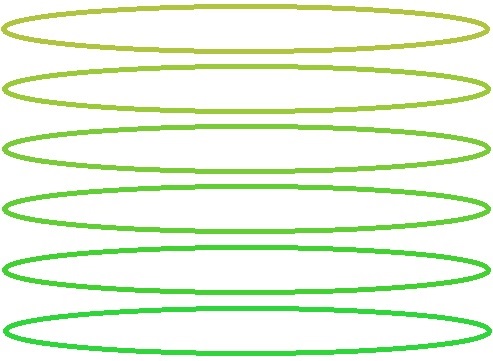}\\
\end{center}
\caption{Perceived Ring Structure of Space}
\end{figure}
\section{Detection}
We first observe that the length of both the nanowires and nanotubes are discrete. The smallest unit of such discrete length  being $2 \alpha \hbar$. We also observe that  both the nanowires and nanotubes are made up of discrete atoms, and GUP deformation occurs by considering the leading order correction to the continuum approximation of this discrete atomic system \cite{gupb}. 
Now as we obtain a  discrete structure due to GUP deformation, we can equate this   discreteness due to GUP with   inter-atomic length scale $l$. Thus, we can write $2\alpha \hbar = l$, and so the exact value of $\alpha$ will depend on the inter-atomic length, which in turn will depend on the  material used to construct such nanowires and nanotubes. For example, for graphene the interatomic length is $l_{gr}= 1.42 \times 10^{-10} m$, so we have $\alpha  = l_{gr}/ 2 \hbar = 7.2 \times 10^{23} {m J^{-1} s^{-1}}$. Similarly, for $TiO_2$, we have $\alpha = l_{TiO_2}/ 2 \hbar = 4.6 \times 10^{-10}/ 2 \hbar = 2.3 \times 10^{24} {m J^{-1} s^{-1}}$. This is still much smaller than the effect that occurs due to the minimal length being of the order of Planck length $\alpha_0 = l_{Pl}/2 \hbar = 8 m J^{-1} s^{-1} $ (where $\alpha_0$ represents the correction due to quantum gravitational effects \cite{ml15}). Thus, the relative effect of such inter-atomic distances compared to Planck length is given by $\alpha_0/\alpha$, which for graphene is $\approx 10^{-23}$ and for $TiO_2$ is $\approx 10^{-24}$. Thus, these effects from next to nearest atom hopping are considerable easier to detect than those which occur due to quantum gravitational effects \cite{ml15}.

Now such discreetness due to GUP will modify the quantum transport in such nanowires and nanotubes. The quantum transport in nanowires and nanotubes can be studied by connecting such nanostructures  to non-collinear ferromagnets via tunnel junctions \cite{1214}. It has already been observed that  the low energy description of such  junctions can be analyzed using   conformally invariant
boundary conditions, which  represent exchange coupling  \cite{1214}. The quantum transport in such a ferromagnet-nanotube-ferromagnet device or a ferromagnet-nanowire-ferromagnet device can be tested by measuring the current-voltage relation for them. 
The GUP deformation of such nanowires and nanotubes would in turn change the quantum transport in such nanostructues. This will modify the the low energy description of such  junctions, if we again couple such nanowires and nanotubes to  non-collinear ferromagnets via tunnel junctions. 
It  will be possible to test the effects produced from  GUP deformation    by measuring the current-voltage relation for  device constructed using such junctions. A GUP deformation will modify such current-voltage relation, and such modification can be detected using current technology. This is because  it  is possible to  accurately measure such current-voltage relation for  nano-scale device using current technology \cite{tech1, tech2}. Thus, the deformation studied in this paper can be tested experimentally.

\section{Conclusion} 
In this paper, we have analyzed the effects of  GUP deformation on nanowires and nanotubes. The GUP deformation of such systems is important as it can  occur as a first order correction to the continuum approximation of a discrete atomic  condensed matter systems. The GUP deformation can also occur due to quantum gravitational effects, and, even in this case, it would be important to analyze the GUP deformation of such nanostructures. As in that case, ultra-precise measurements of these nanostructures would be used to test such quantum gravitational effects. Thus, without specifying the origin of GUP, we have analyzed the effect of GUP deformation on nanowires and nanotubes. As these nanowires and nanotubes have cylindrical topology, we have analyzed the GUP  deformation of  the Schrodinger equation with cylindrical topology.  Then specific solutions to the  deformed Schrodinger equation with different boundary conditions were  used to analyze the effects of GUP deformation on  nanowires and nanotubes. This was used to analyze the effect of GUP deformation on the energy in nanowires and nanotubes. It was observed that due to non-perturbative effects the background geometry of such nanotubes and nanowires would become discrete. We have discussed the detection of such GUP effects using a ferromagnet-nanotube-ferromagnet device or a ferromagnet-nanowire-ferromagnet device.

It would be interesting to analyze such a deformation of nanowires and nanotubes using Dirac materials. Such materials are described by the relativistic Dirac equation. Thus, we can analyze a Dirac equation with cylindrical topology. Then we can GUP deform that Dirac equation. We can use this GUP deformed Dirac equation for investigating the behavior of nanowires and nanotubes made of Dirac materials. It would be also interesting to investigate the quantum transport in such  nanowires and nanotubes made from Dirac materials.  Using the GUP deformed Dirac equation with cylindrical topology, we can investigate the  effect of GUP deformation on the quantum transport properties of these nanowires and nanotubes. 
It is also possible to analyze the effect of  noncommutative deformation of nanowires and nanotubes. It may be noted that just like GUP deformation, there are also two motivations to investigate the noncommutative deformation of nanowires and nanotubes. The noncommutative deformation can occur from external fields in condensed matter systems. It can also occur from specific models of quantum gravity, and then the motivation to study its effects on a condensed matter system would be to use ultra-precise measurements of that condensed matter system to test such effects. 
Finally, the techniques used here can be used to analyze other systems with cylindrical topology.  It would also be interesting to analyze  Landau levels in such nanowires and nanotubes, and the effect of GUP deformation  on the Landau levels in them. This could then be used to analyze the effects of the GUP deformation on quantum Hall effect in these nanostructures. It is expected that the quantum Hall effect will be modified due to the modification of Landau levels from GUP deformation.  It would also be interesting to discuss the possible detection of quantum Hall effect in nanowires and nanotubes. 

\section*{Acknowledgments}
Salman Sajad Wani acknowledges support from the Scientific and Technological Research Council of Türkiye (TÜBİTAK) BİDEB 2232-A program under project number 121C067. 

     {}

\end{document}